\documentclass[twocolumn,aps,prl]{revtex4}

\usepackage{graphicx}
\usepackage{dcolumn}
\usepackage{bm}

\begin{document}

\preprint{APS/123-QED}

\title{Superlattice
Magnetophonon Resonances in Strongly Coupled InAs/GaSb
Superlattices\\}

\author{R.S. Deacon}
\affiliation{Clarendon Laboratory, University of Oxford, Parks Road, Oxford OX1 3PU, United Kingdom.\\
}%
\author{A.B. Henriques}
\affiliation{Instituto de Fisica, Universidade de Sao Paulo, Caixa,
Postal 66318, 05315-970 Sao Paulo, Brazil.}
\author{R.J. Nicholas}%
\email{r.nicholas@physics.ox.ac.uk}
\affiliation{Clarendon Laboratory, University of Oxford, Parks Road, Oxford OX1 3PU, United Kingdom.\\
}%
\author{P.A. Shields}%
\affiliation{Clarendon Laboratory, University of Oxford, Parks Road, Oxford OX1 3PU, United Kingdom.\\
}%

\date{\today}

\begin{abstract}
We report an experimental study of miniband magnetoconduction in
semiconducting InAs/GaSb superlattices. For samples with miniband
widths below the longitudinal optical phonon energy we identify a
new superlattice magnetophonon resonance (SLMPR) caused by
resonant scattering of electrons across the mini-Brillouin zone.
This new resonant feature arises directly from the superlattice
dispersion and total magnetic quantisation (energetic decoupling)
of the superlattice Landau level minibands.
\end{abstract}


\maketitle

Semiconductor superlattices (SL's) comprise alternating layers of
two or more semiconductor materials, leading to the formation of
continuous energy bands in the growth direction called minibands.
The reduced Brillouin zone and energy band width of the SL allows
measurements that probe parameter spaces which are inaccessible in
bulk semiconductors. Total quantisation of the superlattice energy
scheme can be achieved by the application of a large magnetic
field which suppresses inter-Landau level miniband (LLMB)
scattering and allows the realisation of a 'quasi' 1-dimensional
or quantum box SL (QBSL) regime. This has led to strong interest
in the SL magnetoresistance and transport characteristics
\cite{Paper:Fowler,Paper:QuenchMori2004,Paper:Mori2001,Paper:Henriques,
Paper:WSLDeacon1}. The SL miniband structure can be engineered
such that in the QBSL regime optical phonon scattering is
limited\cite{Paper:Noguchi-Sakaki,Paper:Sakaki} by using narrow
minigap and miniband widths, leaving only weak acoustic-phonon
processes to dissipate the electron energy. In this report we
investigate magnetotransport properties of InAs/GaSb superlattices
in the miniband transport regime. In a previous
publication\cite{Paper:WSLDeacon1} we investigated hot-electron
magnetophonon resonance caused by the LO phonon mediated hopping
between Landau Wannier-Stark states at low temperatures. In this
report we study longitudinal magnetophonon resonances caused by
the resonant emission/absorption of longitudinal optical (LO)
phonons in the miniband transport regime for a range of SL
structures at high tempertures. Through a systematic study using
different miniband widths we identify a new form of magnetophonon
resonance which provides evidence for the energetic de-coupling of
SL Landau level minibands leading to suppression of optical phonon
scattering.

\begin{figure*}[t!]
\centering
\includegraphics[width=0.9\linewidth]{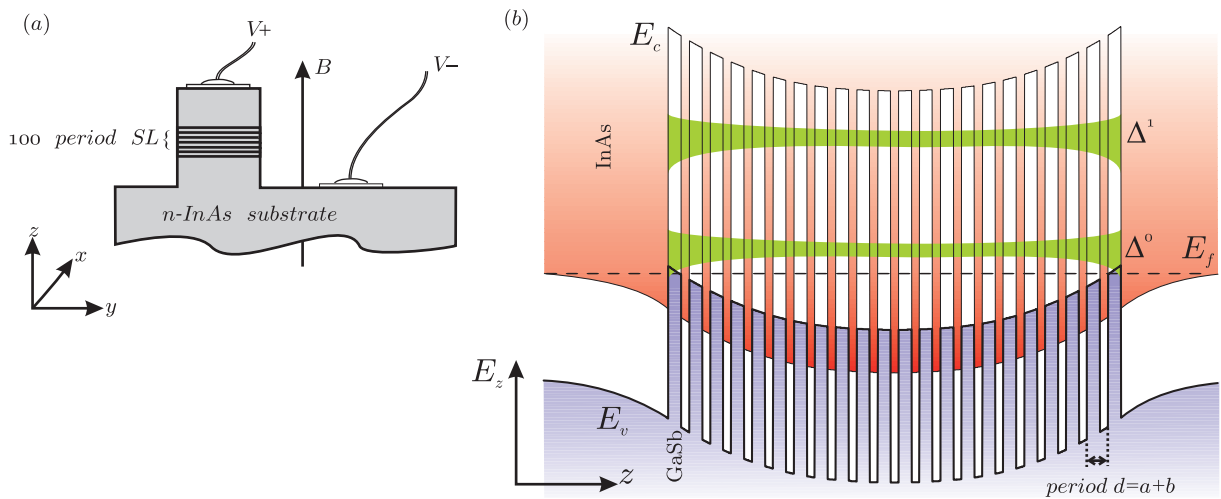}
\caption{(\textit{a}) Schematic of sample geometry. (\textit{b})
Schematic of the sample potential profile for a 22 period SL without
an applied bias displaying the effect of sample-substrate and
sample-cap inversion layers which broaden the superlattice miniband
and facilitate electron injection into the structure. This schematic
picture is supported by self-consistent \textbf{k$\cdot$p}
simulations.\label{fig:Samples-potential}}
\end{figure*}

\section{Introduction}

The Bloch frequency ($\Omega$) of a biased SL is given by
$\Omega=eFd/\hbar$ where $F$ is the applied electric field and $d$ is
the superlattice period. Superlattice transport at low temperatures
is characterised by two regimes. At low electric fields where
$\hbar\Omega<\Delta$, where $\Delta$ is the SL miniband width,
miniband transport through extended SL states dominates. In the
simple Esaki-Tsu miniband transport model\cite{Paper:EsakiTsu}
electron drift velocity ($\nu$) is described by a scattered Bloch
oscillator in 1-dimension, $\nu=\frac{\mu F}{1+(F/F_{c})^{2}}$, where
electron mobility $\mu= e\Delta \tau d^{2}/2\hbar^{2}$,
$F_{c}=\hbar/e\tau d$ and $\tau$ is the scattering time. The main
mechanisms which contribute to $\tau$ are phonon scattering, impact
ionisation and interface roughness scattering. At high electric
fields however $\hbar\Omega>\Delta$ causing the miniband to split
into localised Wannier-Stark-ladder (WSL)
states\cite{Paper:WSQ,Paper:WSLopt,Paper:Kast} and consequently
miniband transport is no longer the dominant process.


Despite the critical role of energy relaxation processes in
superlattices the magnetophonon effect\cite{Paper:Gurevitch} has only
received a small amount of attention. It has been extensively studied
in bulk\cite{Paper:Stradling,Paper:Harper,Paper:RobinReview1} and
2-dimensional\cite{Paper:RobinReview2} semiconductor systems, where
magnetophonon resonances (MPR) are observed as magnetoresistance
oscillations caused by resonant scattering of electrons by optical
phonons. The dominant electron-phonon coupling in all III-V systems
is with the LO phonon due to the large electric polarisation
associated with these modes. Conservation of crystal momentum limits
the LO phonon scattering to phonons near the Brillouin zone center
such that the LO phonon energy is essentially mono-energetic.

The effects of MPR on the resistivity depend strongly on the
relative configuration of the applied electric and magnetic
fields. In the transverse case, ($F\perp B$), resonant momentum
relaxation causes magnetoresistance maxima to be observed at
precisely the MPR resonance condition

\begin{equation}
    \delta n\omega_{c}=\omega_{LO}\ ,
\label{eqn:MPR}
\end{equation}

\noindent where $\omega_{c}$ is the cyclotron frequency,
$\omega_{LO}$ is the LO phonon frequency and $\delta n$ is an
integer. At B-fields which satisfy the MPR condition strong inelastic
scattering occurs between zero momentum states separated in Landau
index by $\delta n$. In the longitudinal configuration ($F\parallel
B$) considered in this report the MPR is typically more complicated
as direct LO phonon emission and absorption processes between zero
momentum states do not relax the electron momentum in the
electric-field direction. In bulk materials resonances occur due to
an interplay of different indirect scattering
processes\cite{Paper:Harper,Paper:Barker}. Extensive experimental
observations of longitudinal MPR (LMPR) in bulk III-V
systems\cite{Paper:Stradling,Paper:Harper} reveal resistivity minima
which are displaced to $B$-fields somewhat below condition
\ref{eqn:MPR}. Calculations for superlattices have suggested both
that resistivity minima\cite{Paper:Kleinert} should occur at fields
slightly below the resonance condition and that maxima could
occur\cite{Paper:Shu92,Paper:PolyanovskiiA} at the resonance
condition. When $\Delta$ is significantly below the LO phonon energy
Polyanovskii\cite{Paper:PolyanovskiiA,Paper:PolyanovskiiB} has also
suggested that transitions from the top to the bottom of the miniband
will generate new superlattice magnetophonon resonances (SLMPR) at
the condition

\begin{equation}
\delta n\hbar\omega_{c}=\hbar\omega_{LO} + \Delta\ .
\label{eqn:Polyanovskii}
\end{equation}

\noindent where resonant LO phonon scattering between areas of
high density of states (DOS) at the top and bottom of the
superlattice miniband enhances the current.

In contrast to the predictions of theory all experimental
observations of MPR in
Superlattices\cite{Paper:MPR1,Paper:MPR2,Paper:MPR3} have assigned
MPR features as resistance maxima, predominantly in samples with
low $\Delta$ and using the GaAs/AlAs system. A significant
enhancement of the resonant peak intensity, compared with bulk
material, has been reported and attributed to the effect of the
superlattice band structure. Noguchi \textit{et
al.}\cite{Paper:MPR1} also observed that low $\Delta$ samples
exhibited plateaus in the oscillatory part of the
magnetoresistance trace which it was suggested may provide
evidence for Polyanovskii's predictions. Detailed studies by
Gassot \textit{et al.}\cite{Paper:MPR2} also assigned LMPR peaks
as maxima in resistance but reported significant deviations from
the predicted MPR conditions. The analysis of the GaAs/AlAs system
is further complicated by the presence of resonances attributed to
both the GaAs and AlAs LO phonons, which are approximately 25\%
different in energy.

This paper reports studies of the magnetophonon effect in the type-II
InAs/GaSb system which has a low carrier effective mass
(m$^{\ast}\sim 0.05$m$_{e}$) allowing the study of transport at
cyclotron energies considerably above the LO phonon energy
($\hbar\omega_{LO}=30$\,meV). The analysis of the resonance is also
simplified because the LO phonon energies of the InAs and GaSb layers
are almost exactly equal. Tunnelling between adjacent superlattice
layers is dominated by interband coupling and is strongly \textbf{k}
dependant. The interband coupling to the barrier valence band states
is strongly reduced for higher Landau index ($n$) LLMB's resulting in
narrower miniband widths\cite{Paper:Hales} ($\Delta_{n}(B)$) and
suppressing domain formation.

Throughout this report we will refer to all experimentally
observed resonances as LMPR features so as to distinguish them
from the MPR condition. This is important as the longitudinal
configuration typically produces resonances slightly shifted from
the MPR condition.

\section{Experimental Method}

\begin{figure}[h]
\centering
\includegraphics[width=1.0\linewidth]{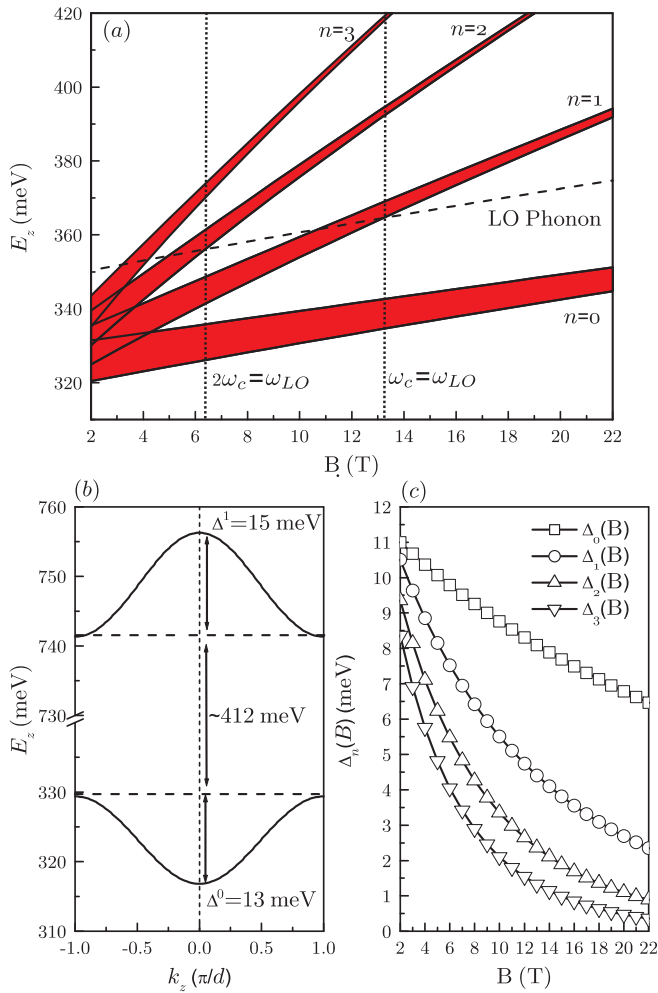}
\caption{Example \textbf{k$\cdot$p} simulation results for sample 4561 (\textit{a}) Landau miniband fan calculated from average of spin up and down levels. The dashed line marks the LO phonon energy and fundamental MPR condition. (\textit{b}) Miniband dispersion for the 1$^{st}$ and 2$^{nd}$ SL minibands at $B=0$\,T. (\textit{c}) Plot of the magnetic field dependence of the first four Landau minibands.\label{fig:4561LL}}
\end{figure}

Experiments were performed on 100 period undoped InAs/GaSb
superlattices grown by MOVPE. Samples were grown on
\textit{n}-type InAs substrates (Carrier density $\sim 5\times
10^{15}$ cm$^{-3}$ estimated from Shubnikov-de Haas measurements)
with $5000$\,\AA\ InAs buffer and cap layers. The ratio of
InAs/GaSb ($a:b$) is estimated from growth rates measured using an
\textit{in-situ} surface photo-absorption (SPA)
technique\cite{Paper:SPA}. Samples were also characterized using
X-ray diffraction (XRD) allowing calculation of the superlattice
period ($d=a+b$).

For vertical transport measurements 150\,$\mu$m mesas were defined
using standard lithographic and wet etching techniques. Ohmic
contacts were made to sample and substrate by evaporating 5\,nm of
chromium and 150\,nm of gold. Previous
measurements\cite{Paper:WSLDeacon1,Paper:WSLDeacon2} on the same
samples have identified that transport is best described by a
number of active periods which is less than the nominal 100
periods grown. Reduction in the number of periods is caused by the
formation of InAs inversion layers at the superlattice-substrate
and superlattice-cap interfaces, figure
\ref{fig:Samples-potential} (\textit{b}). Low temperature
measurements on these samples identify 2-dimensional electron
gases with carrier densities $\sim 1\times 10^{12}$ cm$^{-2}$,
from Shubnikov-de Haas like oscillations, which is consistent with
the formation of this structure. Electron wavefunctions in the
miniband couple with the inversion layer states causing a widening
of the superlattice miniband in some of the first and last periods
of the structure. The broadened miniband states are highly
conducting and allow electrons to be easily injected into the
miniband. The broadened miniband also exhibits negligible voltage
drop when the structure is biased resulting in the reduction of
the number of active periods observed($p$).

All measurements were performed in dc mode with two contacts as
displayed schematically in figure
\ref{fig:Samples-potential}(\textit{a}). $R(B)$ measurements in
magnetic fields up to 19.5\,T were performed using a
superconducting magnet system and a \emph{Keithley} 236 SMU.
Higher magnetic fields were obtained using non-destructive magnets
in the Clarendon laboratory Kurti pulsed magnetic field facility.
The magnetic field pulse was recorded using a pick-up coil
situated 2\,mm below the sample. Induced voltages during the
magnetic field pulse are removed by averaging two pulses of
opposite polarity. Pulsed field measurements were performed with
\emph{Gage Compuscope} 5\,MHz transient recorder cards.

\begin{table}[b]
\centering
\begin{tabular}{c|c|c|c|c}
\hline Sample No.& $d$ (\AA)\footnote[1]{Measured with XRD.} &
$d_{InAs}:d_{GaSb}$\footnote[2]{Estimated with SPA data.} &
$\Delta_{n=0}(0)$ (meV) \footnote[3]{Estimated using
$\mathbf{k\cdot p}$ calculations ($\pm 15$\,\% error).} & $p$\footnote[4]{Estimated from Stark-cyclotron-resonances and hot-electron MPR see references \cite{Paper:WSLDeacon1} and \cite{Paper:WSLDeacon2}.}\\
\hline
4577 & 166 & 0.29 & 2 & 56\\
4562 & 130 & 0.46 & 10 & 60\\
4561 & 126 & 0.50 & 13 & 59\\
4579 & 117 & 0.83 & 27 & 62\\
3756 & 93 & 0.82 & 48 & 67\\
4520 & 86 & 1.26 & 75 & 60\\
\hline
\end{tabular}
\caption{Sample characteristics.\label{tab:growth}}
\end{table}

Sample characteristics were simulated with \textbf{k$\cdot$p}
theory\cite{Paper:kp1} solved using the envelope function
approximation\cite{Paper:BastardEFA} in momentum
space\cite{Paper:RefWebsite}. Sample characteristics are
summarised in table \ref{tab:growth}. Fourier transform
magnetoabsorption spectroscopy was used to observe the
superlattice energy gap which allows refined estimation of the
ratio $a:b$ and shows that miniband width estimates in table
\ref{tab:growth} are correct within to $\pm 15$\,\%.
\textbf{k$\cdot$p} simulation of the superlattice Landau miniband
fan are later used to predict the position of MPR features. An
example of a SL Landau miniband fan is presented in figure
\ref{fig:4561LL} for sample 4561. All samples display a large
separation between the first and second conduction minibands
($\sim 400$\,meV for 4561) such that for the bias values used in
this study only conduction through the fundamental miniband need
be considered, figure \ref{fig:4561LL} (\textit{b}). Simulation
results also clearly display a strong suppression of miniband
width with increasing magnetic field which is stronger for higher
index Landau levels due to the higher energy of these states,
figure \ref{fig:4561LL} (\textit{c}).

\section{Results}

\begin{figure}[t]
\centering
\includegraphics[width=0.9\linewidth]{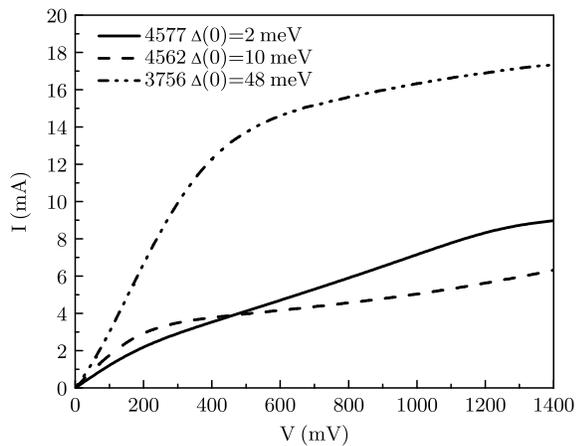}
\caption{$I(V)$ characteristic of sample 4577 and 4562 at $T=80$\,K
and 3756 at $T=77$\,K. \label{fig:I-VEsaki-Tsu}}
\end{figure}

$I(V)$ curves for three samples are displayed in figure
\ref{fig:I-VEsaki-Tsu} and are characteristic of the response of
all superlattices studied in this report. We observe conventional
miniband transport characteristics with a region of ohmic miniband
transport at low bias followed by a miniband transport peak
identified as a shoulder in the $I(V)$ trace (for example at $\sim
200$\,mV for sample 4577 and $\sim 600$\,mV for sample 3756). For
sample 4577 we observe a second shoulder at $\sim 1250$\,mV which
is attributed to electron-phonon resonances in the Stark hopping
regime\cite{Paper:Govorov,Paper:Starkphonon} as have been
discussed in a previous publication on hot-electron MPR observed
in the WSL regime for the same sample set\cite{Paper:WSLDeacon1}.

\begin{figure}[b]
\centering
\includegraphics[width=0.9\linewidth]{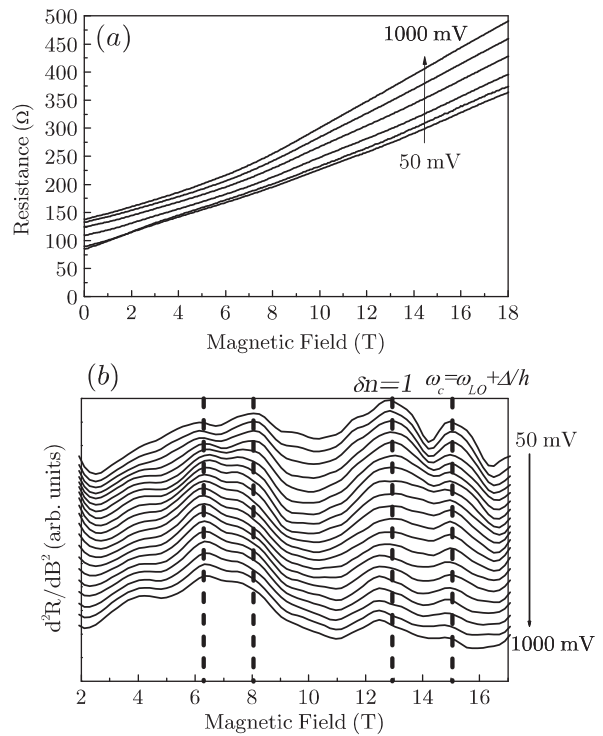}
\caption{(\textit{a}) $R(B)$ curves for sample 4577 at $T=80$\,K
for a range of bias. (\textit{b}) Plot of $d^{2}R/dB^{2}$ for
sample 4577 with $T=80$\,K and a range of sample bias. Vertical
dashed lines mark the observed LMPR and SLMPR features.
\label{fig:4577-RB}}
\end{figure}

$R(B)$ curves for sample 4577 ($\Delta_{0}(0)=2$\,meV) are
displayed in figure \ref{fig:4577-RB} (\textit{a}) for bias in the
range $0-1000$\,mV and $T=80$\,K. $R(B)$ traces are analysed by
plotting $d^{2}R/dB^{2}$ to remove the monotonically increasing
background resistance and identify weak resonances, figure
\ref{fig:4577-RB} (\textit{b}). In plots of $d^{2}R/dB^{2}$ maxima
correspond to resistivity minima in the equivalent $R(B)$ trace.
In this paper we will use the simplification that the additional
resistivity maxima and minima caused by the resonant scattering
processes refer to minima and maxima in $d^{2}R/dB^{2}$
respectively, a simplification in discussion often used when
discussing MPR features\cite{Paper:MPRReview}. Plots of
$d^{2}R/dB^{2}$ display strong maxima features at $B\sim 13,15$\,T
and $B\sim 6.3, 8$\,T. Estimates of the $\delta n=1$ MPR condition
for sample 4577 using $\mathbf{k\cdot p}$ calculations predict a
resonance at $B=14.2$\,T.

The appearance of two resonant peaks close to the conventional
magnetophonon condition predicted by $\mathbf{k\cdot p}$
calculations suggests that the resonant behavior may be
significantly more complex than has previously been assumed. The
main resistivity maximum lies close to the conventional MPR
condition, equation \ref{eqn:MPR}, whilst the resistivity minimum
which might be attributed to LMPR is at $B\sim 12.8$\,T shifted to
$\sim 8\,\%$ below the conventional MPR condition. The second
minimum at $B\sim 15$\,T is a possible example of the new SLMPR
resonance given by equation \ref{eqn:Polyanovskii}. This
assignment is supported by the fact that both LMPR and SLMPR
features are significantly suppressed with increasing bias beyond
$\sim$200 mV when the superlattice moves into the WSL regime, and
so all further measurements were restricted to the miniband
transport range of bias. The feature at $\sim 8$\,T is likely to
be the $\delta n=2$ SLMPR resonance. Assignment of the 6.3T
feature to $\delta n=2$ LMPR is however inconclusive as a residual
Shubnikov de-Haas oscillation peak for the bulk substrate occurs
at approximately $6$\,T and complicates analysis in this low
B-field region.

\begin{figure}[t]
\centering
\includegraphics[width=0.45\textwidth]{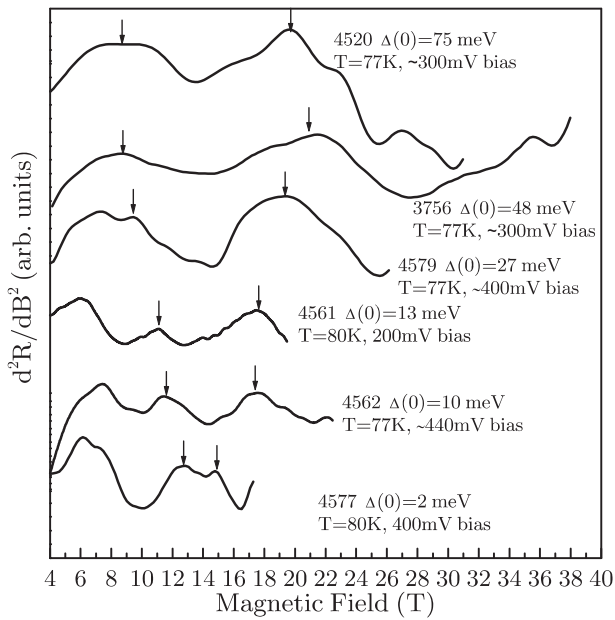}
\caption{Plot of $d^{2}R/dB^{2}$ for a range of superlattice
samples.}
\label{fig:Allplots} 
\end{figure}

In order to test this assignment results are examined for samples
with a range of miniband widths as shown in figure
\ref{fig:Allplots}. The strong features identified for analysis
are selected for repeatability between successive measurements and
in comparison with data taken for steady fields up to $19$\,T, as
shown for sample 4577 in figure \ref{fig:4577-RB}. Resistivity
minima ($d^{2}R/dB^{2}$ maxima) which are candidates for LMPR and
SLMPR features are indicated by arrows. As the miniband width
increases the separation of the SLMPR feature from the
conventional MPR resonance increases progressively. All features
show a characteristic MPR temperature dependence with maximum
amplitude in the range $T=80-140$\,K.


\begin{figure}[t]
\centering
\includegraphics[width=0.45\textwidth]{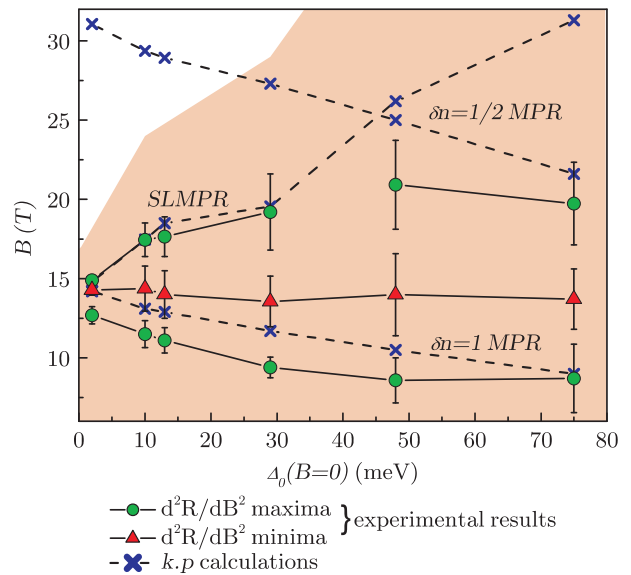}
\caption{Plot of predicted LMPR and SMPR features from
\textbf{k$\cdot$p} calculations alongside experimentally observed
resonances. The x-axis indicates the SL miniband width at $B=0$\,T
calculated with \textbf{k$\cdot$ p}. Note that the
\textbf{k$\cdot$p} calculated resonance positions take account of
the reduction in SL miniband width with increasing magnetic field
which becomes particularly significant for the three highest
miniband width samples. As there is no predicted lineshape for the
resonances the errors are estimated from 40\% full width half
maximum (FWHM) of the features. Shaded region indicates the
parameter space experimentally probed.}
\label{fig:SLMPR} 
\end{figure}

\section{Discussion}

The magnetic field positions of the observed LMPR and SLMPR minima
and maxima features are compared with the values predicted by
$\mathbf{k\cdot p}$ calculations and Equations \ref{eqn:MPR} and
\ref{eqn:Polyanovskii} in figure \ref{fig:SLMPR}. The
$\mathbf{k\cdot p}$ predicted magnetic field value for the MPR
condition falls with increasing $\Delta$ due to the decrease in
the superlattice band gap, which causes the effective mass to
decrease. We observe that resistivity maxima ($d^{2}R/dB^{2}$
minima) occur at fields slightly above the conventional MPR
resonance position. The experimentally observed resistivity minima
labelled as $\delta n=1$ LMPR are typically seen when
$\omega_{c}/\omega_{LO}=0.87\pm 0.06$ in good agreement with bulk
LMPR resonances\cite{Paper:Stradling,Paper:Harper}. By contrast
the SLMPR features move up in field due to the increased
contribution from $\Delta$ in equation \ref{eqn:Polyanovskii} and
show excellent agreement with the predicted SLMPR positions when
account is taken of the magnetic field dependence of the miniband
width.

\begin{figure*}[t]
\includegraphics[width=0.95\textwidth]{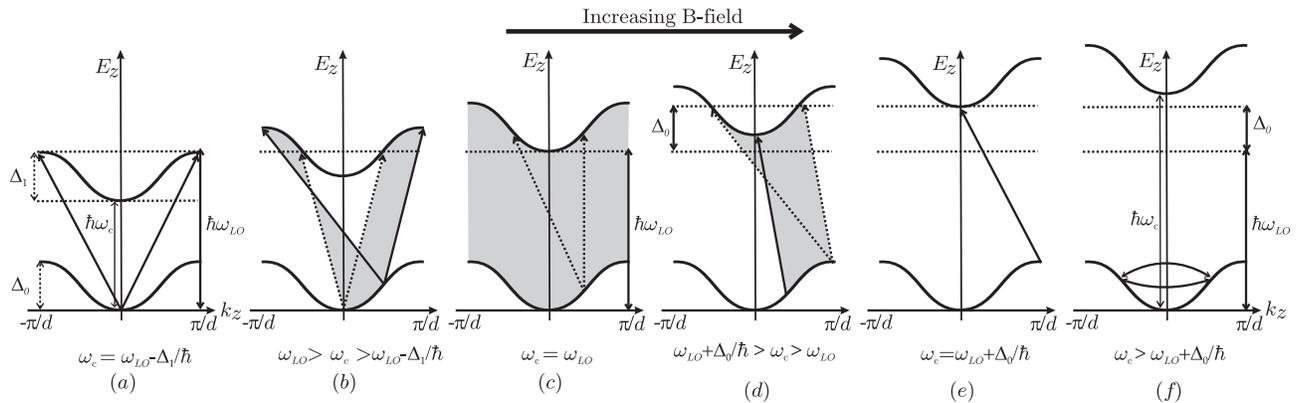}
\caption{Schematics of the LO phonon mediated transitions for narrow
miniband width samples ($\Delta_{0}(0)<\hbar\omega_{LO}$) between the
$n=0$ and $n=1$ LLMB's. Transitions are highlighted at (\textit{a})
$\omega_{c}=\omega_{LO}-\Delta_{1}/\hbar$, (\textit{b})
$\omega_{LO}>\omega_{c}>\omega_{LO}-\Delta_{1}/\hbar$, (\textit{c})
$\omega_{c}=\omega_{LO}$ (The MPR condition), (\textit{d})
$\omega_{LO}+\Delta_{0}/\hbar>\omega_{c}>\omega_{LO}$, (\textit{e})
$\omega_{c}=\omega_{LO}+\Delta_{0}/\hbar$ and (\textit{f})
$\omega_{c}>\omega_{LO}+\Delta_{0}/\hbar$. The allowed LO phonon
mediated transitions from areas of positive $k_{z}$ are indicated by
the shaded region.}
\label{fig:SLMPRv2} 
\end{figure*}

As the miniband width increases we expect a continuous transition
to conventional bulk magnetophonon behavior once
$\Delta_{0}(B)>\hbar\omega_{LO}$ since the superlattices may be
considered to act as a bulk 3-dimensional system if all thermal
and cyclotron excitations are much smaller than the superlattice
miniband width

\begin{equation}
    k_{B}T,\hbar\omega_{c}, \hbar\omega_{LO}\ll \Delta_{0}(B)\ .
\label{eqn:Bulk-SL}
\end{equation}

Within this regime electrons are unable to probe the upper portion of
the superlattice miniband and carriers only experience a dispersion
similar to the parabolic bulk case. It would therefore be expected
that such samples display no SLMPR features. $\mathbf{k\cdot p}$
calculation results for samples 3756 and 4520 indicate that if
observed the SLMPR would be located at $B=26.1$\,T and $B=31.3$\,T
respectively. In experiment (figure \ref{fig:Allplots}) both exhibit
resistance minima features at around $20$\,T which correspond well
with the predicted $\delta n=1/2$ MPR resonances which is also
particularly strong in the Longitudinal configuration for bulk
materials\cite{Paper:Harper} suggesting that the transition to bulk
behavior has occurred for these structures. This will be discussed
further later in this report.

The physical origins of the LMPR and SLMPR features in low miniband
width samples ($\Delta_{0}(0)<\hbar\omega_{LO}$) can be understood by
considering the LO phonon absorption/emission processes which are
allowed for different $\omega_{c}$ values, as shown in figure
\ref{fig:SLMPRv2}. Inter-miniband LO phonon scattering between the
$n=0$ and $n=1$ LLMB's is only permitted in the range
$\omega_{LO}-\Delta_{1}/\hbar<\omega_{c}<\omega_{LO}+\Delta_{0}/\hbar$
due to the forbidden energy gaps in the system. Note also that the
miniband width of any given LLMB has a significant magnetic field
dependence ($\Delta_{n}(B)$) further complicating this schematic
picture. Transitions for which final or initial states are at
$k_{z}=0$ and/or the mini-Brillouin zone boundary $k_{z}=\pm\pi/d$
have high scattering rates due to the large density of states at
these $k_{z}$.

We can consider the scattering of electrons in the positive
$k_{z}$ region of the $n=0$ LLMB dispersion which contribute to
the transport current. It is clear that LO phonon absorption
transitions will result in final states at $+k$ and $-k$ such that
the average velocity of a final state is zero. The average of LO
phonon scattering events therefore relaxes the electron momentum
and the resistance will be proportional to scattering rate. The
scattering rate will increase approaching the MPR condition as LO
phonon scattering is allowed over a greater proportion of the
dispersion. At the MPR condition the LO phonon scattering rate is
at its greatest. If $\Delta_{0}=\Delta_{1}$ scattering will be
allowed anywhere within the superlattice dispersion. In reality
$\Delta_{0}>\Delta_{1}$ for the InAs/GaSb system restricting
scattering at the mini-Brillouin zone boundary.

Above the MPR condition the scattering rate is reduced with
increasing B-field as LO phonon transitions from the center of the
superlattice dispersion are increasingly forbidden. At the SLMPR
condition ($\omega_{c}=\omega_{LO}+\Delta_{0}/\hbar$) only
scattering from $k_{z}=\pm\pi/d$ in the $n=0$ LLMB to $k_{z}=0$ in
the $n=1$ LLMB is allowed. The relaxation therefore links two high
density of states regions of the dispersion relation resulting
once again in a high scattering rate.

\begin{figure*}[t]
\centering
\includegraphics[width=0.7\linewidth]{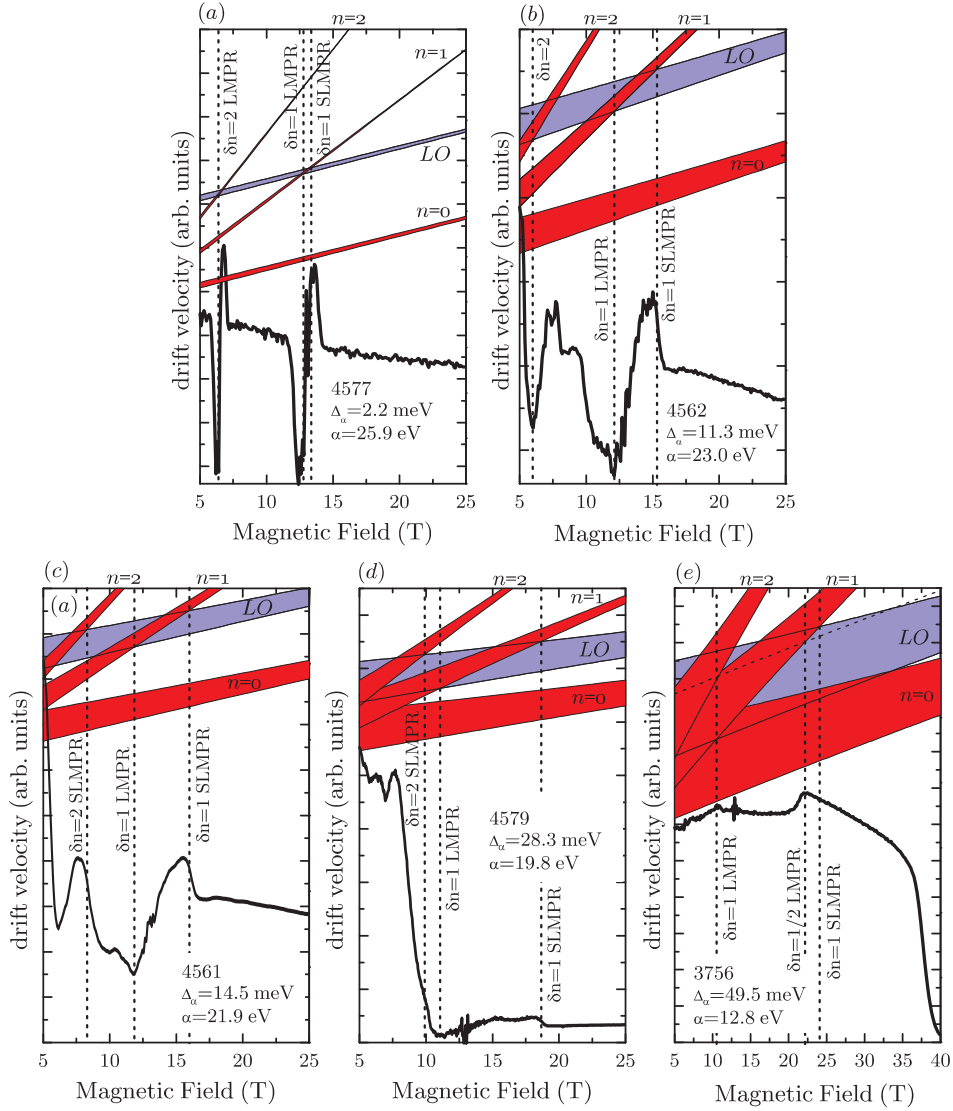}
\caption{Monte Carlo simulation results for (\textit{a}) sample 4577,
$\Delta_{\alpha}=2.2$\,meV, $\alpha=25.9$\,eV$^{-1}$,(\textit{b})
sample 4562, $\Delta_{\alpha}=11.3$\,meV, $\alpha=23.0$\,eV$^{-1}$,
(\textit{c}) sample 4561, $\Delta_{\alpha}=14.5$\,meV,
$\alpha=21.9$\,eV$^{-1}$, (\textit{d}) sample 4579,
$\Delta_{\alpha}=28.3$\,meV, $\alpha=19.8$\,eV$^{-1}$ and
(\textit{e}) sample 3756, $\Delta_{\alpha}=48.5$\,meV,
$\alpha=12.8$\,eV$^{-1}$. All are simulated for $T=100$\,K and
$F=F_{c}/2$. Shaded bands represent the approximated Landau level
fan. Vertical dashed lines mark the SLMPR and MPR conditions. For
sample 3756 a diagonal dashed line indicates the $n=0$ LLMB at
$k_{z}=0$ plus $2\hbar\omega_{LO}$, marking the $\delta n=1/2$ MPR
condition.\label{fig:Andre2}}
\end{figure*}

For superlattice miniband transport, however, the current flow is
dependent on more than simply the immediate scattering rate. The
contribution to current flow also depends on the subsequent
ballistic motion in k-space under electric field acceleration,
since the velocity is strongly dependent upon where the carrier is
in the dispersion relation. These factors must be considered to
fully understand the form of MPR features. We consider the
relative effects of acceleration of carriers in initial and final
scattered states to determine the effect of the scattering upon
transport current. Acceleration in an electric field following LO
phonon scattering has the effect of decreasing drift velocity
below the MPR condition since the carrier velocity is increasing
with $k$ in the lower Landau level and decreasing with $k$ in the
upper level (figure \ref{fig:SLMPRv2}(b)). Conversely the drift
velocity increases above the MPR condition. At the MPR condition
final and initial states are comparable such that subsequent
acceleration does not alter the total transport current.
Scattering at the SLMPR condition has the most dramatic and
significant consequence that it prevents the occurrence of Bragg
scattering and the subsequent suppression of transport current due
to cycling of the carriers through the negative drift velocity
section of the dispersion relation. Above
$\hbar\omega_{c}=\hbar\omega_{LO}+\Delta_{0}(B)$ the coupling of
the $n=0$ and $n=1$ LLMBs through LO phonon scattering is removed
and intra-LLMB scattering dominates. This reverses carrier
momentum and quenches
transport\cite{Paper:QuenchMori2004,Paper:Henriques} resulting in
a significant increase in resistance following the SLMPR feature.
The characteristics of the LMPR and SLMPR features are therefore
determined by interplay of the scattering rate and the subsequent
ballistic motion of the transport electrons. This simple analysis
indicates that LMPR is likely to appear as a resistance maximum at
the MPR condition due to the high scattering rate and SLMPR as a
resistance minimum followed by a large increase in resistance.

In order to understand the different contributions to the
magnetoresistance, modelling of the experimental results was
therefore performed using semiclassical Monte Carlo simulations of
the miniband transport, assuming that the effects of Wannier-Stark
localisation are not significant. Simulations follow the method
outlined by Henriques \emph{et al.}\cite{Paper:Henriques}. Limited
inelastic acoustic phonon scattering in an energy window of 1\,meV
(selected to be less than $\Delta$) was introduced to ensure that the
conductivity at high $B$-fields is small but non-zero. It is found
that changing this window significantly alters the background
magnetoresistance but that the LMPR and SLMPR features remain.
Simulation results are displayed for superlattice band structures
approximating to that of samples 4561, 4579 and 3756 in figure
\ref{fig:Andre2}. The Landau miniband energy widths have been
approximated with the function
$\Delta_{n}(B)=\Delta_{\alpha}e^{-\alpha(n+\frac{1}{2})\hbar\omega_{c}}$
where $\Delta_{\alpha}$ and $\alpha$ are parameters obtained from
fits to the results of $\mathbf{k\cdot p}$ simulations. Both samples
are simulated with $T=100$\,K and electric field $F=F_{c}/2$ such
that results are close to the ohmic miniband transport regime but
show some ballistic behavior.

The simulation results for low miniband widths (figure
\ref{fig:Andre2} (\textit{a}) and (\textit{b})) display two clear
features: firstly, there is a pronounced minimum in the drift
velocity (corresponding to a resistivity maximum) at the conventional
MPR condition and secondly there are peaks in drift velocity at the
SLMPR conditions. These results show excellent agreement with the
schematic picture previously discussed. The simulations identify that
the magnetotransport is dependent on both the scattering rate and the
positions on the dispersion curves of the initial and final states.
Resistance maxima at the MPR are caused by the increased scattering
rate. The resistance minima at the SLMPR are formed from the relative
effects of electric field acceleration in the initial and final
scattering states.

Similar simulation results are observed for other samples which show
that the position of the SLMPR feature is well described by equation
\ref{eqn:Polyanovskii}, shifting progressively away from the LMPR
feature with increasing miniband width. The simulation results change
significantly when the miniband width becomes large
($\Delta_{0}(B)>\hbar\omega_{LO}$) as observed for sample 3756,
figure \ref{fig:Andre2} (\textit{c}). We observe that resistivity
minima now occur at the $\delta n=1$ and $\delta n=1/2$ MPR
conditions. This is in good agreement with the bulk approximation in
which the LMPR is observed as a resistivity minimum slightly below
the MPR condition\cite{Paper:Harper}. For sample 3756 the $\delta
n=1/2$ MPR feature is particularly enhanced by proximity to the SLMPR
condition. The SLMPR feature is however suppressed above the $\delta
n=1/2$ MPR condition due to efficient intra-miniband LO phonon
emissions.

Our overall conclusion therefore is that the modelling provides
strong support for the attribution of the high field resistivity
minimum to the predicted SLMPR feature and that this is strongly
supported by the clear dependence of the SLMPR position on miniband
width.

By contrast, the attribution of a maximum or minimum in resistance to
the precise MPR condition is not found to hold experimentally. The
stronger feature is found to be a maximum at fields a few percent
above the MPR condition. A resistivity minimum is found in all
samples slightly below the MPR condition, in an analogous behavior to
that seen in bulk material. The non-observation of the SLMPR feature
in the earlier GaAlAs based work is probably due to the complications
caused by the presence of the two phonon modes and the relatively
small range of miniband widths studied.

In summary we have demonstrated a new form of magnetophonon resonance
which occurs due to the scattering of carriers from the top of the
ground state miniband to the bottom of the next LLMB. This new
resonance known as superlattice MPR (SLMPR) was first predicted by
Polyanovskii in 1983. The resonance condition corresponds to taking
carriers from the top of the miniband dispersion where they will have
a negative differential velocity and moving them to the bottom of the
band where they can be 'recycled' through the positive conduction
regime.

Parts of this work has been supported by EuroMagNET under the EU
contract RII3-CT-2004-506239 of the 6th Framework 'Structuring the
European Research Area, Research Infrastructures Action' and by the
Brazilian agency CNPq under contract 308116/2004-6.


\begin{thebibliography}{33}
\bibitem[{Pap()}]{Paper:RefWebsite}
\bibinfo{note}{An example of results obtained using this method can be found
  at: www-rjn.physics.ox.ac.uk/kp/}.
\end{thebibliography}
\end{document}